\documentstyle[twocolumn,prl,aps,epsfig]{revtex}
\begin{document} 
\title{A metallic phase in quantum Hall systems  
due to inter-Landau-band mixing} 
\author{Gang Xiong$^{1,2}$, Shi-Dong Wang$^1$, Qian Niu$^3$, 
De-Cheng Tian$^2$ and X. R. Wang$^1$} 
\address{$^1$ Physics Department, The Hong Kong University of 
Science and Technology, Clear Water Bay, Hong Kong SAR, China} 
\address{$^2$ Physics Department, Wuhan University, Wuhan, Hubei 430072,  
China}  
\address{$^3$ Physics Department, The University of Texas at Austin, 
Austin, Texas 78712-1081} 
\address{\mbox{ }}
\address{\parbox{14cm}{\rm \mbox{ }\mbox{ }
The electronic eigenstates of a quantum Hall (QH) system are chiral 
states. Strong inter-Landau-band mixings among these states can occur 
when the bandwidth is comparable to the spacing of two adjacent Landau 
bands. We show that mixing of localized states with {\it opposite 
chirality} can delocalize electronic states. Based on numerical results, 
we propose the existence of a {\it metallic} phase between two adjacent 
QH phases and between a QH phase and the insulating phase. This result 
is {\it consistent} with non-scaling behaviors observed in recent 
experiments on quantum-Hall-liquid-to-insulator transition. 
}}
\address{\mbox{ }}
\address{\parbox{14cm}{\rm PACS numbers: 73.40.Hm, 71.30.+h, 73.20.Jc}}
\maketitle


\vspace{-0.5cm}

Recently there is a great renewal of interest on transitions from 
integer quantum Hall effect (IQHE) states to an insulator\cite
{klz,jiang,shahar,kravchenko,hilke}. According to the scaling 
theory of localization\cite{abrahams}, all electrons in a 
disordered two-dimensional system are localized in the absence of 
a magnetic field. In the presence of a strong magnetic field, 
a series of disorder broadened Landau bands (LBs) will appear, 
and extended states resides at the centers of these bands while 
states at other energies are localized. The integrally quantized 
Hall (QH) plateaus are observed when the Fermi level lies in the 
localized states, with the value of the Hall conductance, 
$\sigma_{xy}=ne^2/h$, related to the number of filled LBs ($n$). 
As a function of the magnetic field, the Hall conductance jumps 
from one QH plateau to another when the Fermi energy crosses an 
extended-state level. 
Many previous studies\cite{klz,jiang,shahar,kravchenko,hilke} 
have been focused on how such a transition occurs. 

One overlooked issue regarding IQHE is the {\it nature} of a 
transition from one QH plateau to another. All existing theories 
assume it to be a continuous quantum phase transition. The 
fingerprint of a continuous phase transition is scaling laws 
around the transition point, and this assumption is mainly due to 
the early scaling experiments\cite{wei}. In the case of IQHE, a 
continuous quantum phase transition means algebraical diverge of 
the longitudinal Hall-resistivity slope in temperature $T$ at the 
transition point. However, recent experiments\cite{hilke} showed 
that such slopes remain finite when it is extrapolated to $T=0$. 
This implies a {\it non-scaling} behavior around a transition 
point, contradicting to the expectation of continuous quantum 
phase transitions suggested by the theories. It also means 
that one should reexamine the nature of plateau transitions. 

In this letter we shall show that a narrow metallic phase may 
exist between two adjacent IQHE phases and between an IQHE phase 
and an insulating phase when the effect of inter-band mixing of 
opposite chirality is taken into account. Thus, it corresponds 
to two consecutive quantum phase transitions instead of one
when the Hall conductance jumps from one plateau to another, 
consistent with the non-scaling behavior observed in experiments.

According to the semiclassical theory\cite{chalker}, an electronic 
state in a strong magnetic field and in a smooth potential can be 
decomposed into a rapid cyclotron motion and a slow drifting 
motion of the guiding center. The kinetic energy of the cyclotron 
motion is quantized by $E_n=(n+1/2)\hbar\omega_c$, where $\omega_c$ 
is the cyclotron frequency and $n$ is the Landau band (LB) index. 
The trajectory of the drifting motion of a guiding center is along 
an equipotential contour of value $V_0=E-E_n$, where $E$ is the 
total energy of the electron. The equipotential contour consists 
of many loops, and each loop corresponds to one quantum state. 
The loops are localized around potential valleys for $V_0<0$ and 
around peaks for $V_0>0$. The drifting direction of each loop is 
{\it unidirectional}, i.e., chiral states. If one views the plane 
from the direction opposite to the magnetic field, the drifting 
is clockwise around valleys and counter-clockwise around peaks. 
In the absence of inter-Landau-band mixing, it has been shown 
that\cite{chalker} extended states only reside at $V_0=0$, 
i.e., at the center of each LB. 

\begin{figure} 
 \vspace{0mm} 
  \vbox to 4.4cm {\vss\hbox to 8.0cm  
  {\hss\   
    {\includegraphics{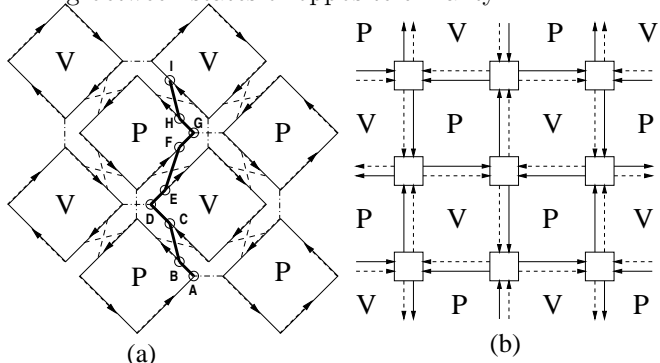} 
    } 
  \hss} 
 \vspace{1mm} 
 } 
\caption{(a) Topological plot of the trajectories of the drifting 
motion of guiding centers (rhombus). The drifting motion around 
a potential peak (valley) is denoted by P (V), and their direction 
are indicated by the arrows. Dashed lines stand for inter-band 
mixing, and doted lines for tunneling at saddle points. The thick 
line (A to I) describes the trajectory of an electron due to a 
strong inter-band mixing. (b) The equivalent two-channel 
Chalker-Coddington network model of (a). Solid and dashed lines 
on each link denote two channels from two LBs. 
Squares stand for saddle points. P, V and arrows have 
the same meaning as those in (a).}
\label{network} 
\vspace{0mm} 
\end{figure}

In a weak magnetic field or strong disorders, the width of the 
Landau subbands may be comparable with the Landau gap, and 
inter-band mixing can no longer be ignored. In order to 
investigate the consequences of this mixing, we shall consider a 
simple system of two adjacent LBs. Our interest will be on the 
localization properties of electronic states between the two bands. 
Using the semiclassical theory described in the previous paragraph, 
two sets of loops are obtained, one for the upper LB and the other 
for the lower LB. The two sets have {\it opposite chirality}, and 
they are spatially separated. The set of loops in the upper band 
is localized around valleys denoted by {\it V} in Fig. \ref
{network}(a) and the other set around peaks denoted by {\it P}. 
Assume that tunneling can occur between two neighboring localized 
states (loops) from the same band near their saddle point and two 
nearby states of different bands are coupled due to impurity 
scattering, Fig. \ref{network}(a) is topologically equivalent to 
a two-channel version of Chalker-Coddington (CC) network 
model\cite{wang} as shown in Fig. \ref{network}(b). In a 
one-channel CC network model, one and only one extended state 
exists in every LB center, corresponding to the percolation point. 
Turn on the inter-band mixing of the {\it opposite chirality}, 
localized loops may become less localized. In order to see this, 
let us consider an extreme case that no tunneling at saddle points, 
but with such a strong inter-band mixing that an electron will 
move from a loop, say around a peak, to a neighboring loop around 
a valley as shown by $B\to C$ in Fig. \ref{network}(a). Following 
an electron starting at A, the trajectory of this electron will 
be $A\to B\to C\to D\to E...$. The electron does not move on a 
closed loop anymore, delocalized! On the other hand, for two sets 
of loops with the {\it same chirality} (that is both of them are 
localized around either valleys or peaks), the inter-band mixing  
mainly occurs between two loops localized around the same 
position, and this mixing cannot help much to delocalize a state. 
This is why we shall consider mixing between spatially separated 
states with opposite chirality, but neglecting the mixing between 
the states of the same chirality separated only in energy. However, 
it does not mean that the mixing of the same chirality has not 
effect at all. As it was found in some previous works\cite{wang}, 
this kind of mixing may shift an extended state from its LB center. 
Level shifting due to mixing between states of the same
chirality may distore the shape of the phase diagram, but should not
alter its topology. The merging of the bands of extended states is
exclusively due to the mixing between states of opposite chirality.

Electron localization length is often obtained from the transfer  
matrix method. For a 2-dimensional system, however, it is well known 
that this quantity is difficult to provide a conclusive answer to 
questions related to the metal-insulator transition (MIT)\cite{xie}.  
On the other hand, level-statistics analysis\cite{klesse} has been 
used in studying MIT. We follow the approach proposed by Klesse 
and Metzler\cite{klesse}. A quantum state of the network model can 
be expressed by a vector $\vec{\Phi}=(\{\phi_i^u,\phi_i^l\})$, where 
$\phi_i^u$ and $\phi_i^l$ are the wavefunction amplitudes of the 
$i$-th link of the upper band ({\it u}) and the lower band ({\it 
l}), respectively. As shown by Fertig\cite{fertig}, the network 
model can be described by an {\it evolution operator} $U(E)$ which 
is an $E$-dependent matrix determined by the scattering properties 
of nodes and links in the model. The eigen-value equation of this 
operator is $U(E)\vec{\Phi}_{\alpha}(E)=e^{i\omega_{\alpha}(E)}\vec
{\Phi}_{\alpha}(E)$, where $\alpha$ is the eigenstate index. The 
eigen-energies $\{E_n\}$ of the system are those $E$ with $\omega$ 
being an integer multiples of $2\pi$. It was shown\cite{klesse} 
that the level-spacing statistics of the set of {\it quasi-energies
} $\{\omega_{\alpha}(E)\}$ is the same as that of the set of $\{E_n
\}$ near $E$. Therefore, the localization property of states with 
an energy $E$ can be obtained. A great advantage of this approach 
is that all the eigenvalues of the evolution operator $U(E)$ can be 
used in the analysis. A characteristic quantity $I_0$, defined by 
$I_0=\int s^2P(s)ds/2$, is used to examine the localization 
property, where $P(s)$ is the level-spacing distribution function 
of the quasi-energies\cite{klesse}. It is well-known that $I_0=1$ 
for localized states\cite{klesse}. Thus, we use the following 
simple criteria. If $I_0$ of a state with energy $E$ increases and 
approaches $1$ with lattice size $L$, this states is localized. 
Otherwise, it is extended. We studied the model for L=8, 12, 16 
(and 20 for some energy values) with periodic boundaries along both 
directions. For each energy value $E$, a sufficient number of 
ensembles are used in order to collect more than $5\times10^4$ 
data points in our analysis. 

In the following analysis, we assume that inter-band mixing takes 
place only on the links. Tunneling at each node occurs for the 
same band and is described by a SO(4) matrix 
\begin{equation} 
    S= 
    \left ( 
    \begin{array}{llll}  
    s_u^R & s_u^L & 0 & 0 \\ 
    -s_u^L & s_u^R & 0 & 0 \\ 
    0 & 0 & s_l^R & s_l^L \\ 
    0 & 0 & -s_l^L & s_l^R 
    \end{array}
    \right ), 
\end{equation} 
where the subscripts $u$ stands for the upper band and $l$ for the 
lower LBs. The elements $s_{u(l)}^L$ and $s_{u(l)}^R$ are related 
to the scattering probabilities of an incoming wavefunction in the 
upper (lower) band to outgoing channels of its left-hand-side and 
its right-hand-side, respectively. We choose $s_{u(l)}^R=\sqrt
{1-(s_{u(l)}^L)^2}$ because of the orthogonality of the matrix. 
The potential around a saddle point is assumed to be 
$V(x,y)=-Ux^2+Uy^2+V_c$\cite{wang}, 
where $U$ is a constant describing the potential fluctuation and 
$V_c$ is the potential barrier at this point.
The left-hand scattering amplitude is given by\cite{fertig}
\begin{equation} 
s_{u(l)}^L=[1+\exp(-\pi\epsilon_{u(l)})]^{-1/2},
\end{equation}
where $\epsilon_{u(l)}=[E+V_c-(n_{u(l)}+1/2)E_2]/E_1$, $E$ the 
electronic energy, $E_1=\frac{\hbar\omega_c}{2\sqrt{2}}\sqrt{K-1}$ 
and $E_2=\frac{\hbar\omega_c}{\sqrt{2}}\sqrt{K+1}$ with
$K=\sqrt{\frac{64U^2}{m^2\omega_c^4}+1}$. The kinetic energies of 
cyclotron motion in the two bands are $(n_u+1/2)E_2$ and $(n_l+1/2)
E_2$, respectively, where $n_{u(l)}$ are the band indices and 
$\Delta n=n_u-n_l=1$. The dimensionless ratio $E_r=E_2/E_1=2\sqrt
{1+\frac{2}{K-1}}$ approaches $2$ from above as $U$ or the inverse 
of $\omega_c$ increase\cite{fertig}, i.e., the strong-disorder or 
weak-field regime. Since this is the regime that we are interested 
in, we choose the value of it to be 2.2 in our calculations. 
For convenience, we choose $E_2$ as the energy unit 
and the cyclotron energy of the lower band as the reference point. 
The energy range between the two band centers is thus $E\in[0,1]$. 
The potential $V_c$ at a saddle point is treated as a random number 
uniformly distributed within the range of $[-E_1,E_1]$.

Inter-band mixing on a link is described by a U(2) matrix with 
random Aharonov-Bohm phases $\phi_i(i=1,2,3,4)$ accumulated 
along propagation paths, 
\begin{equation} 
    M= 
    \left ( 
    \begin{array}{ll} e^{i\phi_1} & 0 \\ 0 & e^{i\phi_2} 
    \end{array} 
    \right ) 
    \left ( 
    \begin{array}{ll} \cos\theta & \sin\theta \\ -\sin\theta & 
    \cos\theta \end{array} 
    \right )  
    \left ( 
    \begin{array}{ll} e^{i\phi_3} & 0 \\ 0 & e^{i\phi_4} 
    \end{array} 
    \right ), 
\end{equation} 
where $\sin\theta$ describes the mixing strength. 
In our calculations, we shall assume that they are uniformly 
distributed in $[0,2\pi]$\cite{chalker}. $P$, defined as $\sqrt{P/
(1+P)}=\sin\theta$, is used to characterize the mixing strength. 
$P$ will take the same value for all links in our calculations. 
\begin{figure}
 \vspace{0mm} 
  \vbox to 4.9cm {\vss\hbox to 8.0cm 
  {\hss\ 
    {\includegraphics{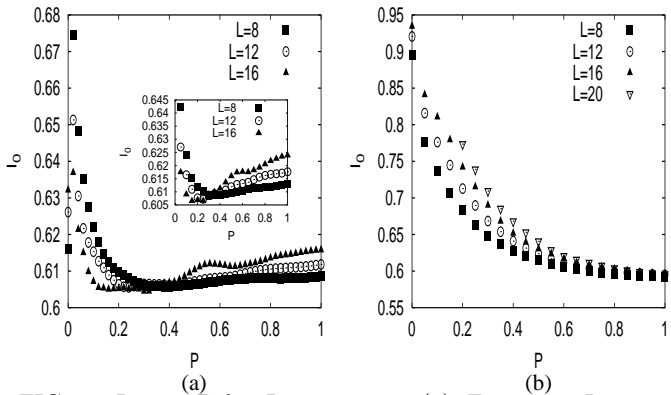} 
    } 
  \hss} 
 } 
 \vspace{0mm} 
\caption{$I_0$ vs. $P$ for $L=8,12,16$, (a) $E=0.02$; 
Inset: $E=0$. (b) $E=0.5$.}
\label{data}   
\end{figure} 
Curves in Fig. 2 are $I_0$ vs. mixing strength $P$ at $E=0.02$ (a); 
0. (inset of (a)); and 0.5 (b) for $L=8,$ 12, 16, 20. The state of 
$E=0.02$ is localized at zero mixing since $I_0$ increases with 
sample size $L$. Fig. 2(a) shows two crossing point at small $P$. 
Between the two points, $I_0$ decreases with sample size $L$. 
Thus, according to the criteria explained before, the state is 
extended in this regime. For the lower band center $E=0$, $I_0$ 
decreases initially with $L$ at small $P$, indicating an extended 
state. Then $I_0$ of different $L$ cross at a particular $P_c$. 
$I_0$ decreases with $L$ for $P>P_c$, showing the feature of a 
localized state. Fig. 2(b) shows that state of $E=0.5$ is 
localized at small $P$ because $I_0$ increases with sample size 
and approaches $1$. For large $P(>1)$, all curves of of different 
system size tend to merge together. 
The independence of $I_0$ on system size suggests a 
localization-delocalization transition at the merging point.

The existence of new extended states at $E\sim0.5$ in the case of
strong inter-band mixing can be understood as follows. Assume that 
the intra-band tunneling at nodes are negligibly weak for states 
of $E\sim0.5$, we saw already from Fig. \ref{network}(a) that the 
maximal inter-band mixing ($\sin\theta=1$) delocalizes the state, 
which is localized at zero inter-band mixing. If one views $p=
\sin^2\theta$ as connection probability of two neighboring loops 
of opposite chirality, our two-channel model without intra-band 
tunnelings at nodes is analogous to a bond-percolation problem. 
It is well-known that a percolation cluster exists at $p\ge 
p_c=1/2$ or $P\ge P_c=1$ for a square lattice\cite{stauff}. 
Therefore, an extended state is formed by strong mixing. 
One hopes that the intra-band tunnelings at nodes will only 
modify the threshold value of the mixing strength. 
\begin{figure} 
 \vspace{0mm} 
   \vbox to 4.2cm {\vss\hbox to 8.0cm 
   {\hss\ 
      {\includegraphics{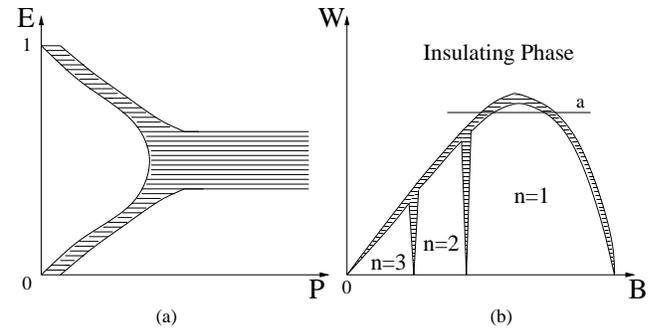} 
      } 
   \hss} 
   } 
   \vspace{1mm} 
\caption{(a) Topological phase diagram of electron localization in 
$E-P$ plane. The shadowed regime is for extended states (metallic 
phase). (b) Topological QH phase diagram in $W-B$ plane. $W$ stands 
for the disorder strength, and $B$ for magnetic field. The shadowed 
regime is for metallic phase. The area indicated by symbol $n$ is 
the $n$-plateau IQHE phase. The rest area is the insulating phase.}
\label{phase} 
\end{figure} 
To express our numerical results in $E-P$ plane, a topological phase 
diagram shown in Fig. \ref{phase}(a) is obtained. In the absence of 
inter-band mixing, only the singular energy level of each LB is 
extended. In the presence of inter-band mixing of opposite chirality, 
there are two regimes. At weak mixing when the inter-Landau-band 
separation is much smaller than the Landau-bandwidth, each of the 
extended state broadens to a narrow band of extended states near the 
LB centers. The extended states in the lowest LB shift from the LB 
centre. These extended states will eventually merge with those from 
the higher LBs. This shifting of extended states was also observed 
before\cite{klz}. At strong mixing, a band of extended states 
exists between two Landau bands where all states are localized 
without the mixing. Systems in a weak magnetic field should 
correspond to the strong mixing case. In terms of the QH plateau 
transitions, a direct transition occurs because two QH phases 
merges together to form a narrow metallic phase in a weak field. 
Thus, we propose that a direct transition from an IQHE phase to the 
insulating phase is realized by passing through a metallic phase.

Plot above results in the plane of disorder and magnetic field, we 
obtain a new topological QH phase diagram as shown in Fig.
\ref{phase}(b). This is similar to the empiric diagram obtained 
experimentally in reference 4. The origin ($W=0, \ B=0$) is singular 
point. According to the weak localization theory\cite{abrahams}, no 
extended state exists at this point. Differing from existing theories, 
there exists a narrow metallic phase between two adjacent IQHE phases 
and between an IQHE phase and an insulating phase. This new phase 
diagram is consistent with the non-scaling experiments\cite{hilke} 
where samples are relatively dirty, and inter-band mixing is strong, 
corresponding to a process along line a in Fig. \ref{phase}(b). 
The system undergoes two quantum phase transitions each time when it 
moves from the QH insulating phase to IQHE phase of $n=1$ and back 
to the Anderson insulating phase as the magnetic field decreases. 
To verify this claim, we analyzed the original experimental data in 
Ref. 6 according to the two quantum phase transition point assumption. 
The standard one-parameter scaling analysis is performed, and good 
scaling behaviors are obtained for two close critical filling 
factors of $\nu_{c1}=0.6453$ and $\nu_{c2}=0.6477$. The critical 
exponents in the left and right-hand sides of the transition 
regime are equal and the value is $z\nu=2.33\pm0.01$. This value 
is consistent with previous works\cite{wei,chalker,wang}.
On the other hand, if we are forced to 
fit the data by one critical point in each transition, we found either 
two distinct values for critical exponents on the two sides of the 
critical point, or the data does not follow the scaling law at all. 
Our fitting shows that the width of the metallic phase regime 
is about $5\times10^{-3} tesla$ while the value of the magnetic 
field was increased by $1\sim2\times10^{-3} tesla$ each time in the 
experiments. This may explain why the metallic phase was overlooked. 

It is worth noticing that two metallic states have been studied
extensively in the QH system. One is the composite Fermion state 
at the half-filling in the lowest Landau level (LL) and the other 
is the stripe state at the half-filled higher LLs. These states 
are formed by the Coulomb interaction effect in the high mobility 
samples. They are different from our metallic phase due to level 
mixing in the paper. Although we have not considered the 
electron-electron interactions in our study, there is no reason 
why delocalization effect due to the level mixing will be destroied 
by the Coulomb interaction. Of course, the interaction could 
cause the level mixing effect with different dependence on the 
magnetic field. Our model can be used to describe spin polarized 
systems. In this case, the two LBs are for spin up and spin down 
states, respectively. Indeed, two-channel CC models have been used 
before to simulate spin resolved problem\cite{wang}. 
These studies 
cannot distinguish an extended state from a localized state between 
the energy region of the two extended states of the two Landau bands.
In this sense, our results are consistent with those of early works. 

In summary, we argued that inter-Landau-band mixing of opposite chirality
can lead to delocalization. Within the network model, we found numerical 
evidences for a metallic phase between two neighboring QH phases 
by using the level-spacing statistics analysis. This new phase diagram 
gives a possible explanation for non-scaling behavior observed in recent 
experiments. The results of a scaling analysis on experimental data 
support our conclusion. We also point out the difference 
between our result and Wang et. al.'s. We have only considered two 
LBs in this work, and we believe that metallic phases exist in higher 
IQHE phases. But, a further study including more LBs is needed 
to answer this question definitely. 

This work was supported by the Research Grant Council of HKSAR, China. 
QN would like to acknowledge the support of NSF-DMR9705406, the 
Welch Foundation and China-NSF. G.X. thanks Dr. Dan Shahar for 
providing the original experimental data.


\vskip -0.8cm
\begin{references} 
\bibitem{klz} S. Kivelson, D. H. Lee and S. C. Zhang, 
Phys. Rev. B {\bf 46}, 2223(1992); D. Z. Liu, X. C. Xie and 
Q. Niu, Phys. Rev. Lett. {\bf 76}, 975(1996); 
D. N. Sheng and Z. Y. Wang, Phys. Rev. Lett. {\bf 80}, 
580(1998); 
X. R. Wang, X. C. Xie, Q. Niu, and J. Jain, cond-mat/0008411.
\bibitem{jiang} H. W. Jiang, C. E. Johnson, K. L. Wang, S. T. 
Hannahs, Phys. Rev. Lett. {\bf 71}, 1439(1993); I. Glozman, C. E. 
Johnson, and H. W. Jiang, Phys. Rev. Lett. {\bf 74}, 594(1995).\ 
\bibitem{shahar} D. Shahar, D. C. Tsui, J. E. Cunningham, 
Phys. Rev. B {\bf 52}, R14372(1995); 
S.-H. Song, D. Shahar, D. C. Tsui, Y. H. Xie,  
Don Monroe, Phys. Rev. Lett. {\bf 78}, 2200(1997).\ 
\bibitem{kravchenko} S. V. Kravchenko, W. E. Mason, J. E. Fureaux,
V. M. Pudalov, Phys. Rev. Lett. {\bf 75}, 910(1995).\
\bibitem{hilke} M. Hilke {\it et. al.}, Phy. Rev. B {\bf 56}, 
15545(1997); D. Shahar {\it et. al.}, 
Solid State Commun. {\bf107}, 19(1998).\ 
\bibitem{abrahams} E. Abrahams {\it et. al.}, 
Phys. Rev. Lett. {\bf 42}, 673 (1979). 
\bibitem{wei} H. P. Wei {\it et. al.}, 
Phy. Rev. Lett. {\bf 61}, 1294 (1988).
\bibitem{chalker} J. T. Chalker and P. D. Coddington, 
 J.Phys. C: Solid State Phys.{\bf 21}, 2665(1988); 
A. G. Galstyan and M. E. Raikh, 
 Phys. Rev. B {\bf 56}, 1422(1997).
\bibitem{wang}  Z. Q. Wang, D. H. Lee and X. G. Wen, Phys. Rev. Lett. 
{\bf 72}, 2454(1994); D. K. K. Lee and J. T. Chalker,
Phys. Rev. Lett. {\bf 72}, 1510(1994);
V. Kagalovsky, B. Horovitz, and Y. Avishai, 
Phys. Rev. B {\bf 52}, 17044(1995).\ 
\bibitem{xie} X. C. Xie, X. R. Wang, and D. Z. Liu, Phys. Rev. Lett.
{\bf 80}, 3563(1998).
\bibitem{klesse} R. Klesse and M. Metzler, Phys. Rev. Lett. 
{\bf 79}, 721(1997); M. Metzler and I. Varga, J. Phys. 
Soc. Jpn. {\bf 67}, 1856(1998); M. L. Mehta, 
`{\it Random Matrices}' , 2nd ed. (Academic Press, 1991).\ 
\bibitem{fertig} H. A. Fertig and B. I. Halperin, 
Phys. Rev. B {\bf 36}, 7969(1987); H. A. Fertig, 
Phys. Rev. B {\bf 38}, 996(1988).\ 
\bibitem{stauff} D. Stauff and A. Aharony, 
 `{\it Introduction to Percolation Theory}' 
 (Taylor and Francis, London, 1994).\
\end{references}
\end{document}